\documentstyle[aps,prl,graphicx,twocolumn]{revtex}
\newcommand{\Exp}[1]{\mathrm{e}^{\mbox{\footnotesize$#1$}}}

\begin{document}

\draft

\title{Nonclassical Interference Effects In The Radiation From Coherently Driven Uncorrelated Atoms}

\author{C. Skornia$^{\ast\dagger}$, J.von Zanthier$^\ast$, G.S. Agarwal$^{\ast\ddagger}$, E. Werner$^{\dagger}$, H. Walther$^\ast$}

\address{$^\ast$ Max-Planck-Institut f{\"u}r Quantenoptik and Sektion  Physik der LMU M{\"u}nchen,  D-85748 Garching, Germany\protect\\  $^\dagger$ Institut f{\"u}r Theoretische Physik, Universit{\"a}t Regensburg, D-93040 Regensburg, Germany\protect\\  $^\ddagger$ Physical Research Laboratory, Navrangpura, Ahmedabad-380 009, India}

\date{\today}

\maketitle

\begin{abstract} 

We demonstrate the  existence  of new nonclassical  correlations in  the   radiation of two atoms, which are coherently driven by a continuous laser source. The photon-photon-correlations of the fluorescence light show a spatial interferene pattern not present in a classical treatment. A feature of the new phenomenon is, that bunched and antibunched light is emitted in different spatial directions. The calculations are performed analytically. It is pointed out, that the correlations are induced by state reduction due to the measurement process when the detection of the photons does not distinguish between the atoms. It is interesting to note, that the phenomena show up even without any interatomic interaction.

\end{abstract}
\pacs{PACS numbers: 42.50.Dv, 42.50.Ar, 03.65.Bz}
\narrowtext

%%%%%%%%%%%%%%%%%%%%%%%%%%%%%%%%%%%%%%%%%%%%%%%%%%%%%%%%%%%%%%%%%%%%%%%%%%%%%%%%%%%%%%%%%%%%%%%%%%%%%%%%%%%%%%%%%%%%%%%%%%%%%%%%%%%%%%%%%%%%%%%%%%%%%%%%%%%%%%%%%%%%%%%%%%%%%%%%%%%%%%%%%%%%%%%%%%%%%%%%%%%%%%%%%%%%%%%%%%%%%%%%%%%%%%%%%%%%%%%%

Resonance fluorescence from a single atom  driven by a coherent field  was the first example allowing to observe nonclassical   effects such as  antibunching   and subpoissonian   statistics \cite{mandel77,mandel83,cresser82}.  These experiments first done with atoms in a beam  have later also been performed using single trapped ions \cite{walther87,toschek92}. Squeezing in resonance fluorescence has also been investigated \cite{thomas98}.
Since the  early work of  Mollow  \cite{mollow69}, Carmichael  and Walls  \cite{walls76}  one has studied at  length the quantum   statistical  characteristics of    radiation   produced by  a   cooperative  system  of   two  and more  atoms \cite{agarwal77,richter91}. Some evidence of cooperative effects have been demonstrated \cite{brewer96}. There are also some recent proposals to investigate cooperative effects including the interactions between the atoms \cite{dalton95,beige}.\\

\begin{figure}[htbp]  
\begin{center}    
\includegraphics[height=4.2cm]{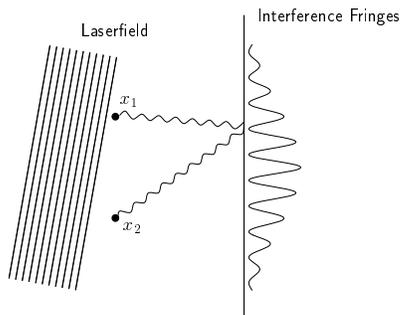}    
\caption{Two atoms in a coherent laser field show interference fringes}
\label{fig:eraser}  
\end{center}
\end{figure}

In the present paper we study the two atom fluorescence leading to new nonclassical effects. We consider a situation, where the two atoms are driven coherently by a continuous laser source. These nonclassical effects are observed in photon-photon-correlations when state reduction occurs in the measurement process. We note, that in an ealier study by Mandel \cite{Mandel83} correlation effects in a similar system have been discussed. However, in contrast to our paper there no continuously pumped atoms have been considered. We calculate nonclassical two photon correlations for all times including dissipation of the atoms. The dynamics resulting from the excitation by cw fields and by spontaneous emission is very important for the present results. We should mention that on the experimental side first order interference in the radiation produced by a prepared system of two atoms has been reported \cite{raizen93}. We also point out, that the creation of entangled states of distant atoms has been discussed in a different context \cite{Cabrillo99}.

It may be noted, that the trapped ion technology is fairly well advanced to envisage resonance fluorescence measurements
with chains of trapped ions driven by coherent fields. One can also use single molecules in crystalline hosts, where
recently nonclassical photon statistics have been measured \cite{orrit99}.\\

%%%%%%%%%%%%%%%%%%%%%%%%%%%%%%%%%%%%%%%%%%%%%%%%%%%%%%%%%%%%%%%%%%%%%%%%%%%%%%%%%%%%%%%%%%%%%%%%%%%%%%%%%%%%%%%%%%%%%%%%%%%%%%%%%%%%%%%%%%%%%%%%%%%%%%%%%%%%%%%%%%%%%%%%%%%%%%%%%%%%%%%%%%%%%%%%%%%%%%%%%%%%%%%%%%%%%%%%%%%%%%%%%%%%%%%%%%%%%%%%

We consider two identical atoms with the  levels $|e\rangle_i$ and  $|g\rangle_i$ ($i=1,2$) at fixed positions ${\mathbf  x}_1$ and ${\mathbf  x}_2$ with dipole moment  $\mathbf{d}$ and transition frequency $\omega$. They are driven  by a resonant external laser field with wavevector ${\mathbf k}_L$.  We assume that the only dissipative terms are due to the spontaneous decays  of the levels $|e\rangle_{1,2}$.  In  the  rotating-wave, Markov  and Born approximations, the time evolution of the system is given by the master equation \cite{agarwal74}

\begin{eqnarray}
\label{eq:MasterEq}
\dot{\rho}&=&    -{\mathrm i} \Omega \sum_{\mu=1}^{2} \left[\Exp{\left[ {\mathrm i} \left( { \mathbf k}_L \cdot {\mathbf x}_{\mu} - \omega_L t \right) \right]} \sigma_\mu^+    + \mbox{h.c.},\rho \right] \nonumber \\  
&&    - \sum_{\mu =1}^{2} \gamma \left(\sigma_\mu^+ \sigma_\mu^- \rho   + \rho \sigma_\mu^+ \sigma_\mu^- - 2 \sigma_\mu^- \rho \sigma_\mu^+ \right),
\end{eqnarray}

where $2  \Omega$ is  the Rabi  frequency of the  atom laser
system, $\sigma^{\pm}_\mu$ are the atomic raising and lowering  operators for atom number  $\mu$, and $2\gamma = \frac43
|{\mathbf  d}|^2 \frac {\omega^3}{\hbar c^3}$  is the  Einstein A coefficient   for the single atom.    As we assume the
electromagnetic field outside the laser beam in the vacuum state, the correlation functions can be written as

\begin{eqnarray} 
\label{eq:G1Def}  
G^{(1)}({\mathbf r},t)  &&=  \sum_{\mu,\nu=1}^2 \Exp{\left[ {\mathrm i}k {\mathbf x}_{\mu,\nu}\cdot\hat{{\mathbf r}}\right]}  \langle \sigma_{\mu}^+\left(t \right) \sigma_{\nu}^-\left(t \right) \rangle\\  \label{eq:G2Def}
G^{(2)}({\mathbf r}_1,t;{\mathbf r}_2,t+\tau)&&=   \sum_{\mu,\nu,\lambda,\rho=1}^2   \Exp {\left[ {\mathrm i}k ({\mathbf x}_{\lambda,\nu}\cdot\hat{{\mathbf r}}_{1}+    {\mathbf x}_{\rho,\mu}\cdot\hat{{\mathbf r}}_{2})\right]} \times \nonumber\\  
&&\langle \sigma_{\lambda}^+\left(t \right) \sigma_{\rho}^+\left(t +\tau\right)  \sigma_{\mu}^-\left(t +\tau\right) \sigma_{\nu}^-\left(t \right) \rangle 
\end{eqnarray}

with ${\mathbf x}_{\mu,\nu} := {\mathbf x}_{\mu}-{\mathbf x}_{\nu}$ and $\hat{{\mathbf r}}=\frac{\mathbf r}{|{\mathbf r}|}$.\\
The two time correlations (\ref{eq:G2Def}) can be calculated using the quantum regression theorem and the time evolution of the density operator.  Therefore the normalized second order correlation function can be written as

\begin{equation}  
\label{eq:g2}  
g^{(2)}({\mathbf r}_1,t;{\mathbf r}_2,t+\tau) =  \frac{P({\mathbf r}_2,t+\tau|{{\mathbf r}_1,t})}{P({\mathbf r}_2,t)}
\end{equation}

where $P({\mathbf r},t)$ is the probability of finding a photon at  position ${\mathbf r}$  at time $t$, and $P({\mathbf  r}_2,t+\tau|{{\mathbf r}_1,t})$ the conditional  probability of finding a photon  at position ${\mathbf r}_2$  at time $t+\tau$ premising the detection of a photon at point ${\mathbf r}_1$ at time t.  Using the inequality

\begin{equation}  
\label{eq:ineq_class_1}  
\mathrm{Tr}(\rho(\alpha I({\mathbf r}_1)+\beta I({\mathbf r}_2))^2)\geq 0 \quad \forall \alpha,\beta 
\end{equation}

leads us to

\begin{equation} 
\label{eq:ineq_class_2}  
\prod_{i=1}^{2}(g^{(2)}({\mathbf r}_i;{\mathbf r}_i)-1)\geq
 (g^{(2)}({\mathbf r}_1;{\mathbf r}_2)-1)^2
\end{equation}

for classical systems.  This gives us a good possibility to estimate the nonclassical behavior of our system.\\

%%%%%%%%%%%%%%%%%%%%%%%%%%%%%%%%%%%%%%%%%%%%%%%%%%%%%%%%%%%%%%%%%%%%%%%%%%%%%%%%%%%%%%%%%%%%%%%%%%%%%%%%%%%%%%%%%%%%%%%%%%%%%%%%%%%%%%%%%%%%%%%%%%%%%%%%%%%%%%%%%%%%%%%%%%%%%%%%%%%%%%%%%%%%%%%%%%%%%%%%%%%%%%%%%%%%%%%%%%%%%%%%%%%%%%%%%%%%%%%%

As there can always be found one state which does not interact with the laser field it is suggestive to use this state to
build up the basis for the further calculations. We call this state $|a\rangle$ and define

\begin{eqnarray} 
\label{eq:dicke}  
|e\rangle &:=& |e,e\rangle \nonumber \\  
|s\rangle &:=& \frac{1}{\sqrt{2}}\left(\Exp{-\mathrm{i} \phi}|e,g\rangle+\Exp{\mathrm{i} \phi}|g,e\rangle\right)\nonumber \\  
|a\rangle &:=& \frac{1}{\sqrt{2}}\left(\Exp{-\mathrm{i} \phi}|e,g\rangle-\Exp{\mathrm{i} \phi}|g,e\rangle\right)\nonumber \\  
|g\rangle &:=& |g,g\rangle \\  
\mbox{with} \quad |i,j\rangle & := & |i\rangle_1 \otimes |j\rangle_2   ,\quad \phi := \frac12 {\mathbf k}_{L}\cdot{\mathbf x}_{12}\nonumber
\end{eqnarray}

\begin{figure}[htbp]   
\begin{center}    
\includegraphics[height=4cm]{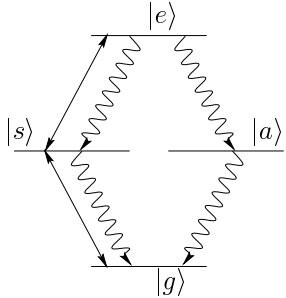}    
\caption{Level scheme of the atomic system for the symmetrized set of states}    \label{fig:levelscheme}   
\end{center}
\end{figure}

In this  representation the master equation reduces to the following sets of 6+3 equations

\begin{eqnarray}  
\label{eq:master}
\dot{\rho}_{ee}&=& 4 ( \alpha \rho_{es}^i - \rho_{ee})\nonumber\\ 
\dot{\rho}_{ss}&=&   2 (\rho_{ee} - \rho_{ss} + 2 \alpha (\rho_{sg}^i- \rho_{es}^i) )\nonumber\\ 
\dot{\rho}_{aa}&=& 2 (\rho_{ee} - \rho_{aa})\nonumber\\ 
\dot{\rho}_{es}^i&=& -3 \rho_{es}^i - 2 \alpha (\rho_{ee} - \rho_{ss} + \rho_{eg}^r )\nonumber\\ 
\dot{\rho}_{sg}^i&=& 2 \rho_{es}^i - \rho_{sg}^i + 2 \alpha (1- \rho_{ee} - \rho_{aa}- 2 \rho_{ss} + \rho_{eg}^r  )\nonumber\\ 
\dot{\rho}_{eg}^r&=& -2 (\rho_{eg}^r + \alpha (\rho_{sg}^i - \rho_{es}^i))\\[1.1ex]
\dot{\rho}_{ea}^r&=&-3 \rho_{ea}^r - 2 \alpha \rho_{sa}^i\nonumber\\
\dot{\rho}_{sa}^i&=&2 (\alpha (\rho_{ea}^r + \rho_{ag}^r) -\rho_{sa}^i )\nonumber\\
\dot{\rho}_{ag}^r&=&-2 \rho_{ea}^r - 2 \alpha \rho_{sa}^i - \rho_{ag}^r
\end{eqnarray}

with $\alpha := \frac{\Omega}{\sqrt{2}\gamma}$, $\rho_{kl}^r:=\Re(\rho_{kl})$ and $\rho_{kl}^i:=\Im(\rho_{kl})$.

By solving the first set of equations (\ref{eq:master}) we find the diagonal elements in the steady-state to be:

\begin{eqnarray} 
\label{eq:Stats}  
\rho_{gg}^{SS} & = & \frac{\left(\gamma^2 + \Omega^2\right)^2}{\left(\gamma^2 + 2 \Omega^2\right)^2}\\  
\rho_{ss}^{SS} & = & \frac{\Omega^2\left[2 \gamma^2+\Omega^2\right]}
  {\left(\gamma^2 + 2 \Omega^2\right)^2} \\  
\rho_{aa}^{SS} =  \rho_{ee}^{SS} &= &\frac{\Omega^4}{\left(\gamma^2 + 2 \Omega^2\right)^2}
\end{eqnarray}

This calculation does not depend on the direction of the driving laser, although it is included by the proper definition of our symmetric and antisymmetric state.

\begin{figure}[htbp]
\begin{center}    
\includegraphics[height=4cm]{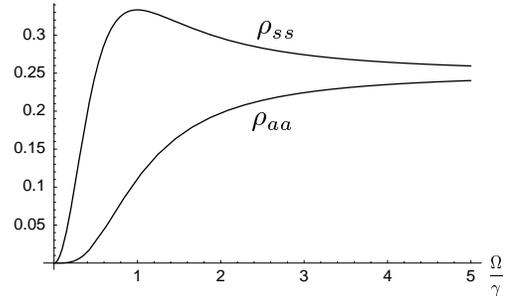}    
\caption{Steady state populations of the symmetric and antisymmetric state as a function of the laser intensity $\Omega$}    
\label{fig:steadystates}  
\end{center}
\end{figure}

If the laser direction is perpendicular to the ion separation ($\phi=0$) this solution corresponds  to the earlier found solutions of Richter \cite{richter91}, where we neglect the dipole-dipole interaction.  This is justified when the ion distance is in the range  of several wavelengths  of the atomic transition. In  this case the solutions are  completely independent of the separation of the atoms.  For strong laser fields ($\Omega\gg\gamma$) the populations are equal for each of
the atomic states ($\rho_{gg}^{SS}=\rho_{ss}^{SS}= \rho_{ee}^{SS}=\rho_{aa}^{SS}=1/4$).\\

%%%%%%%%%%%%%%%%%%%%%%%%%%%%%%%%%%%%%%%%%%%%%%%%%%%%%%%%%%%%%%%%%%%%%%%%%%%%%%%%%%%%%%%%%%%%%%%%%%%%%%%%%%%%%%%%%%%%%%%%%%%%%%%%%%%%%%%%%%%%%%%%%%%%%%%%%%%%%%%%%%%%%%%%%%%%%%%%%%%%%%%%%%%%%%%%%%%%%%%%%%%%%%%%%%%%%%%%%%%%%%%%%%%%%%%%%%%%%%%%

We next discuss, how a detection event leads to state reduction and entanglement.
After a detection at the point ${\mathbf r}$ we have to find a new density operator in the following way

\begin{eqnarray} 
\label{eq:reduction}  
\rho ({\mathbf r})&=&\frac{\sigma^-({\mathbf r}) \rho^{SS}\sigma^+({\mathbf r})}{\langle\sigma^+({\mathbf r})\sigma^-({\mathbf r})\rangle},\quad \mbox{where}\\
\sigma^{\pm}({\mathbf r})&:=&\sigma_1^\pm+\Exp{\mathrm{i}k_L{\mathbf r}{\mathbf      x}_{12}}\sigma_2^\pm.
\end{eqnarray}

If, for instance, the system is in state $|e\rangle$ before a detection at ${\mathbf r}$ with $\hat{{\mathbf r}}{\mathbf x}_{12}=0$, we find it in $|s\rangle$ after the detection. This explains, how entanglement could come into the system, as the  system is transformed from  a non-entangled state $|e\rangle$  to an entangled  state $|s\rangle$ by a detection
event.\\

Let us now check, if the atoms  remain uncorrelated after a  detection event i.e. $\rho({\mathbf  r})$ factorizes into a product of density matrices $\rho_i$ with $\rho_i$ referring to the density matrix  of the i-th atom.  We find that this
is not the case, for example

\begin{equation}  
\label{eq:1sain}
\Im(\rho({\mathbf r})_{sa})=\frac{\Omega^2\sin\phi}{2(\Omega^2+\gamma^2(1+\cos\delta({\mathbf r})))}, \rho_{ee}({\mathbf r})=0
\end{equation}

with  $\delta({\mathbf  r})=({\mathbf k}_{L}  -  k_L  \hat{{\mathbf  r}})\cdot{\mathbf x}_{12}$.  On  the  other  hand a factorized density  matrix would imply,  $\Im(\rho({\mathbf r})_{sa})=0$  after any  detection event.   Furthermore, the dependence of $\Im(\rho({\mathbf  r})_{sa})$  on the coordinates  of the  two atoms  through $\delta({\mathbf r})$  also points out  the nonexistence of a factorized  density matrix after  detection. Thus there are values of ${\mathbf  r}$, where the state reduction by the measurement leaves the system in an entangled state. With this mechanism the entanglement, which is necessary for nonclassical effects, is brought into the system simply by the detection of a single photon.\\

%%%%%%%%%%%%%%%%%%%%%%%%%%%%%%%%%%%%%%%%%%%%%%%%%%%%%%%%%%%%%%%%%%%%%%%%%%%%%%%%%%%%%%%%%%%%%%%%%%%%%%%%%%%%%%%%%%%%%%%%%%%%%%%%%%%%%%%%%%%%%%%%%%%%%%%%%%%%%%%%%%%%%%%%%%%%%%%%%%%%%%%%%%%%%%%%%%%%%%%%%%%%%%%%%%%%%%%%%%%%%%%%%%%%%%%%%%%%%%%%

To calculate the intensity at position ${\mathbf r}$ we use the relations

\begin{eqnarray}  
\label{eq:oprels}  
\langle \sigma_1^+\sigma_1^- \rangle = \langle \sigma_2^+\sigma_2^- \rangle  & = & \rho_{ee}+\frac12\left(\rho_{ss}+\rho_{aa}\right) \\  
\langle \sigma_1^+\sigma_2^- \rangle = \langle \sigma_2^+\sigma_1^- \rangle^*  & = & \frac12\Exp{2\mathrm{i}\phi}\left(\rho_{ss}-\rho_{aa}\right),
\end{eqnarray}

where we assumed the system to be in the steady state (i.e. $\rho_{sa}^{SS}=0$).  We then find

\begin{eqnarray} 
\label{eq:intensity}    
G^{(1)}({\mathbf r},t) & = & \left(1+\cos\delta({\mathbf r})\right)    \left[\rho_{ee}(t)+\rho_{ss}(t)\right] + \nonumber \\ && \left(1-\cos\delta({\mathbf r})\right)    \left[\rho_{ee}(t)+\rho_{aa}(t)\right].
\end{eqnarray}

This leads to the well known interference fringes in the first order correlation function as studied in the experiment of Eichmann et al. \cite{raizen93}.\\

In this case the system  behaves as the well known  double slit experiment  where the atoms act as  slits and the two optical paths interfere.   Like in a  system with prepared  initial states the  photons emitted by symmetric transitions $|e\rangle \rightarrow |s\rangle$ and $|s\rangle  \rightarrow |g\rangle$ show  inverse fringes compared to those emitted by antisymmetric transitions $|e\rangle  \rightarrow |a\rangle$ or $|a\rangle  \rightarrow |g\rangle$. This follows from the  $1\pm\cos\delta({\mathbf r})$ terms in (\ref{eq:intensity}).  One major difference is that  the contrast depends on the intensity of the laser light so that the fringes disappear for higher laser intensities when $\Omega\gg\gamma$. The populations $\rho_{ss}^{SS}$ and $\rho_{aa}^{SS}$ equalize.\\

To get the second order correlation function we solve the master equation by calculating the  Liouville Operator and use for the initial state the density matrix  given by (\ref{eq:reduction}). This procedure  is also equivalent to using the quantum regression theorem.\\

Remarkably enough we are able to give an analytical expression for the intensity-intensity correlation:

\begin{eqnarray}   
\label{eq:2}
&&g^{(2)}({\mathbf r}_1,0;{\mathbf r}_2,\tau) = 1+\frac{e^{-3t}}{4 \nu^2 (s + \cos\delta_1) (s +  \cos\delta_2 )} \times \nonumber\\
&&\big\{4 \Exp{2 t} \nu^2 s \sin\delta_1  \sin\delta_2 + s \left[\Exp{t} \nu^2 s  + (s - 1)^2 \right] \cos\delta_1 \cos\delta_2  \nonumber\\
&&- \Exp{3 t/2} \nu s^2 \left[2 \nu \cos (\nu t) + 3 \sin (\nu t)\right]\nonumber\\
&&+2 \Exp{3 t/2} \nu s   \left(\cos\delta_1  + \cos\delta_2\right) \left[(2s - 3) \sin (\nu t) - 2 \nu \cos (\nu t) \right] \nonumber\\
&&+ \Exp{t/2} \nu \left(2 \Exp{t} \cos\delta_1 \cos\delta_2 + s \sin\delta_1 \sin\delta_2\right)\nonumber\\
&&\qquad\left[2 \nu (s - 2) \cos (\nu t) + (5s - 6)\sin (\nu t) \right] \nonumber\\
&&+\frac14 \cos\delta_1 \cos\delta_2\left[\left(s(s(4s-33)+64)-36\right)\cos (2\nu t)\right.  \nonumber\\
&&\left.+ 2\nu (s - 2) (5s - 6) \sin (2\nu t)\right]\big\}
\end{eqnarray}

with $s: = 2 \left(\frac{\Omega}{\gamma}\right)^2+1$, $\delta_i=\delta({\mathbf r}_i)$ and  $\nu: = \frac12 \sqrt{8s-9}$ being parameters for   the intensity of  the  driving field  and $t:  = \gamma\tau$ is   the time scaled with  the atoms spontaneous emission rate. For $\tau=0$ Eq. (\ref{eq:2}) reduces to

\begin{equation}  
\label{eq:g2zero}  
g^{(2)}({\mathbf r}_1,0;{\mathbf r}_2,0)=  \frac{s^2\cos^2\left(\frac{\delta_1-\delta_2}2\right)}{(s+\cos\delta_1)(s+\cos\delta_2)}
\end{equation}

\begin{figure}[htbp]   
\begin{center}    
\includegraphics[height=5.5cm]{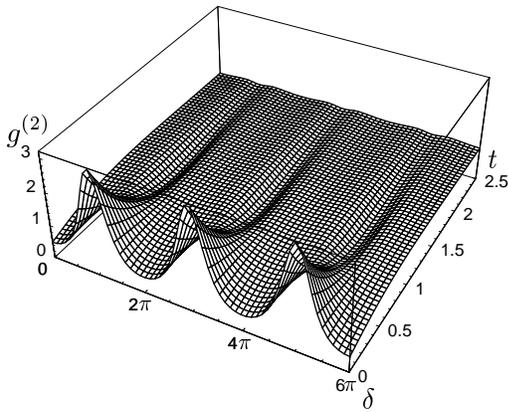}    
\caption{Second order correlation function for one detector with $\delta_1=\delta_2=\delta$, $\Omega=0.8\gamma$}    
\label{fig:intfring}   
\end{center}
\end{figure}

Assuming measurements  with only  one detector at position $\mathbf{r}$, $g^{(2)}({\mathbf r},0;{\mathbf  r},0)$ shows interferences fringes with maxima  larger and minima  smaller than one   (Fig~\ref{fig:intfring}).  This means  that the system is emitting super-poissonian light in one and sub-poissionian light in the other  direction.  This behavior has no analogy in the first order correlation i.e. in the two slit experiment.

\begin{figure}[htbp] 
\begin{center}    
\includegraphics[height=5.5cm]{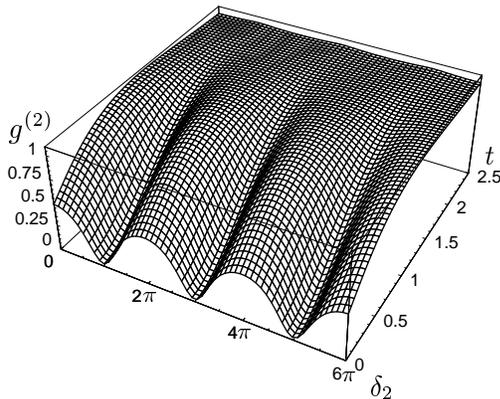}    
\caption{Second order correlation function for two detectors with $\delta_1=2n\pi$, $\Omega=0.8\gamma$}    
\label{fig:d1zerofring}   
\end{center}
\end{figure}

\begin{figure}[htbp]   
\begin{center}    
\includegraphics[height=5.5cm]{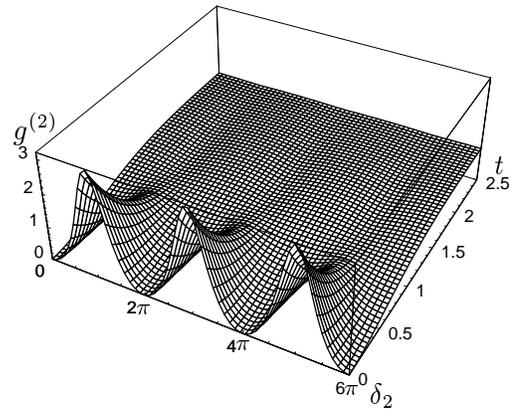}    
\caption{Second order correlation function for two detectors with $\delta_1=(2n+1)\pi$, $\Omega=0.8\gamma$}    
\label{fig:d1pifring}   
\end{center}
\end{figure}

For   two  detectors  the   situation is  different.   For any   detector   position with $\delta_2=(2n+1)\pi+\delta_1$, $g^{(2)}({\mathbf r}_1,0;{\mathbf  r}_2,0)$ vanishes  completely, while at  other  positions  we find  maximum detection probability (Fig~\ref{fig:d1zerofring}). After a detection at position ${\mathbf r}_1$ with $\delta_1=2n\pi$ we  find $\rho_{ee}=\rho_{aa}=0$. Therefore there is no probability to detect any photon at a position  with $\delta_2=(2n+1)\pi$ at the same  time, as no emission into the
antisymmetric channel could take place (\ref{eq:intensity}). So we again find fringes as a function of $\delta_2$. For $\delta_1=(2n+1)\pi$  we  detect an emission on the antisymmetric channel first, so  we  find $g^{(2)}({\mathbf r}_1,0;{\mathbf r}_2,0)$ vanishing for positions ${\mathbf r}_2$  with $\delta_2=2n\pi$ for  the same reasons as above (Fig~\ref{fig:d1pifring}).

After any detection on a specific channel the system needs time for re-excitation to  emit to the orthogonal channel; on the other hand cascade emissions could only take place at the same channel. Note that the behavior of $g^{(2)}({\mathbf r}_1,0;{\mathbf r}_2,\tau)$ is well understood in terms of our symmetrized basis~(\ref{eq:dicke}).  These results show also that the {\bf inequality} (\ref{eq:ineq_class_2}) could be violated by choosing the two detector positions in a way that $g^{(2)}({\mathbf r}_1,0;{\mathbf r}_1,0)>1$  and $g^{(2)}({\mathbf r}_2,0;{\mathbf  r}_2,0)<1$. In this manner we
find nonclassical behavior in our system as consequence of the detection induced entanglement of the two atoms. We note that dependence of the type $\cos^2$ in the two photon correlations has also been derived by Mandel in his work on radiation from atoms prepared in atomic coherent states \cite{Mandel83}. In contrast to that we are dealing here with the more general case of continuous excitation of the atoms including dissipation.\\

%%%%%%%%%%%%%%%%%%%%%%%%%%%%%%%%%%%%%%%%%%%%%%%%%%%%%%%%%%%%%%%%%%%%%%%%%%%%%%%%%%%%%%%%%%%%%%%%%%%%%%%%%%%%%%%%%%%%%%%%%%%%%%%%%%%%%%%%%%%%%%%%%%%%%%%%%%%%%%%%%%%%%%%%%%%%%%%%%%%%%%%%%%%%%%%%%%%%%%%%%%%%%%%%%%%%%%%%%%%%%%%%%%%%%%%%%%%%%%%%

In conclusion it has been shown by analytic calculations,  that a system of two atoms coherently excited by a cw laser source shows nonclassical features being dependent on the distance of the atoms, although there is no assumed interaction between the atoms. This  can be  understood  by measurement induced  entanglement, which comes into the system if  the detection  does not distinguish  between  the  two atoms, i.e. the detected photon does not carry  which-way-information. On the first glance the system studied seems to be similar to the two slit experiment, however the nonclassical effects discussed here do not show up in the first order correlation and can only be seen in the second order correlation as it should be the case for a quantum phenomenon.\\

To see the reported effects single ions stored in a linear rf-trap can be  used; remaining micromotion could be overcome by a phase-sensitive detection  \cite{hoeffges97}. In this paper the case was discussed, that the detector/or detectors do not discriminate the light emitted by the individual atoms. There are, however, further interesting phenomena observable when the which-way-information is available. One example is that the excitation of one atom and the selective observation of the fluorescence from the other one opens, among other effects, the possibility to investigate the dipole-dipole interaction between the atoms. These results will be described elsewhere.

%\bibliographystyle{/home/cks/u210/TeX/bibtex/pra}

%\bibliography{bunch}

\end{document}